\documentclass[twocolumn]{aastex62}
\graphicspath{{./}{figures/}}

\received{2019 September 9}
\revised{2020 March 2}
\accepted{2020 *** **}
\submitjournal{AJ}

\shorttitle{High-resolution optical spectroscopic observations of comet 21P/GZ}
\shortauthors{Shinnaka et al.}

\begin{document}

\title{
\large\bf{
High-resolution optical spectroscopic observations of comet 
21P/Giacobini-Zinner in its 2018 apparition
}
}

\correspondingauthor{Yoshiharu Shinnaka}
\email{yoshiharu.shinnaka@cc.kyoto-su.ac.jp}

\author[0000-0003-4490-9307]{Yoshiharu Shinnaka}
\affil{Laboratory of Infrared High-resolution Spectroscopy (LiH), Koyama Astronomical Observatory, Kyoto Sangyo University, Motoyama, Kamigamo, Kita-ku, Kyoto 603-8555, Japan}

\author[0000-0003-2011-9159]{Hideyo Kawakita}
\affiliation{Laboratory of Infrared High-resolution Spectroscopy (LiH), Koyama Astronomical Observatory, Kyoto Sangyo University, Motoyama, Kamigamo, Kita-ku, Kyoto 603-8555, Japan}
\affiliation{Department of Physics, Faculty of Science, Kyoto Sangyo University, Motoyama, Kamigamo, Kita-ku, Kyoto 603-8555, Japan}

\author[0000-0001-8813-9338]{Akito Tajitsu}
\affiliation{Subaru Telescope, National Astronomical Observatory of Japan, 650 North A'ohoku Place, Hilo, HI 96720, USA}



\begin{abstract}

   Comet 21P/Giacobini-Zinner is a peculiar comet from the viewpoints 
of the chemical and physical properties of its dust grains. We conduct optical 
high-resolution spectroscopic observations of the comet. The intensity 
ratios of forbidden oxygen lines (at 557.7, 630.0, and 636.4 nm) and 
ortho-to-para abundance ratios (OPRs) of water cations (H$_2$O$^+$) and 
amidogen radicals (NH$_2$) are obtained while only the upper limit for 
$^{14}$N/$^{15}$N in the amidogen radical is restricted. The OPRs of H$_2$O$^+$ 
and NH$_2$ are similar to those of other comets, although the real meaning of these 
OPRs is still debated. 
Based on the observation of the forbidden emission lines of oxygen atoms, 
it can be concluded that the comet is depleted in CO$_2$. In consideration with 
the depletion in other highly volatile species found in the near-infrared region and 
the presence of complex organics in comet 21P/Giacobini-Zinner, this comet might form
in a warmer region in the solar nebula compared with other comets.

\end{abstract}



\section{Introduction} \label{sec:intro}
   Comet 21P/Giacobini-Zinner (hereinafter, 21P/GZ) is classified
as a Jupiter-family comet based on its Tisserand parameter with 
respect to Jupiter ($T_{\rm J} = 2.46$). According to the previous reports 
of the observations of comet 21P/GZ, this comet showed unique 
properties of not only volatiles but also dust grains compared with 
other comets: 
(1) depletion of carbon-chain molecules like C$_{2}$ and C$_{3}$
    as well as NH$_{2}$ \citep{Fink2009}, 
(2) depletion of highly volatile species (C$_{2}$H$_{6}$, CH$_{3}$OH, 
    and CO; \citealt{DiSanti2013, DelloRusso2016}), 
and 
(3) negative linear polarization gradient for reflected sunlight 
    by cometary grains, indicative of existence of organic matter 
    \citep{Kiselev2000}. 

   Comet 21P/GZ is also known as the parent comet of the October 
Draconids meteor shower (historically called as the Giacobinids), 
based on the similarity in orbital elements between comet 21P/GZ 
and the meteoroids of the Giacobinids \citep{Jenniskens2006}. 
The meteoroids are thought to be porous grain conglomerates. Their 
derived densities are smaller compared with other meteor showers 
(0.1 -- 0.5 g cm$^{-3}$), and they have typical chondritic abundance 
ratios of the major heavy elements (namely Mg, Fe, and Na), and 
the Giacobinids meteors exhibit fragmentation behaviors 
\citep{Borovicka2010, Borovicka2014} not frequently seen in other 
meteor showers. 

\begin{deluxetable*}{lccccccl}
\tablecaption{Observational circumstances. \label{tab:tab1}}
\tablenum{1}
\tablehead{
\multicolumn{2}{l}{UT Time in 2018} & \colhead{$T_{\rm exp}$ (s)}
& \colhead{$r_{\rm H}$ (au)} & \colhead{$\Delta$ (au)} 
& \colhead{$\dot\Delta$ (km s$^{-1}$)} & \colhead{Airmass} 
& \colhead{Reference stars (Airmass)}
}
\startdata
Sep  5 & 11:27 & 12,300 & 1.015 & 0.396 & $-$2.93 & 2.65--1.18 
       & HD 27026 (1.21), HD 41161 (2.39)\\
Sep  9 & 12:38 &  9,600 & 1.013 & 0.392 & $-$0.48 & 1.74--1.14 
       & HD 27026 (1.08), HD 41161 (2.13) \\
Oct  3 & 13:35 &  6,200 & 1.066 & 0.469 & 10.49 & 1.43--1.18 
       & HD 49643 (3.37), HR 1544 (1.42) \\
\enddata
\tablecomments{The first column indicates the start time of
  the exposures for comet 21P/GZ. $T_{\rm exp}$ is total integration
  time in seconds. $r_{\rm H}$ and $\Delta$ are heliocentric and
  geocentric distances at the observations in au, respectively. 
  $\dot\Delta$ is the relative velocity of the comet to the Earth 
  at the time of observations.}
\end{deluxetable*}

In summary, from the observational viewpoints, comet 21P/GZ is 
peculiar among observed comets. The unique properties
of comet 21P/GZ might be explained by the different birth place 
of the comet (formed under different physical conditions such as 
temperature, dust-to-gas ratio, and ionization degree). 
Therefore, to understand the physical conditions where 
icy/dust materials in comet 21P/GZ formed, we conducted optical 
high-resolution spectroscopic observations of the comet in its 
2018 apparition. We tried to determine those properties of volatiles 
considered as primordial. We report the results of our observations 
and discuss the origin of comet 21P/GZ.

\section{Observations and data reduction} \label{sec:obs}

High-resolution optical spectroscopic observations of comet 21P/GZ 
were performed on UT 2018 September 5, 9, and October 3 using the 
High Dispersion Spectrograph (HDS; \citealt{Noguchi2002}) attached 
to the Subaru Telescope in Maunakea, Hawaii. The heliocentric and 
geocentric distances of the comet were 1.01--1.07 au and 0.39--0.47 
au, respectively. The optical peak of the coma was centered on the 
slit. The spectra covered the wavelength region between 551.4 and 
827.9 nm with a gap between 684.5 and 693.4 nm. The slit size was 
0''.5 $\times$ 8''.5 in the sky. The spectral resolution, $R \equiv 
\lambda/\Delta\lambda$, was 72000 over the entire wavelength region. 
Details of our observations are listed in Table \ref{tab:tab1}.
   Data taken with the Subaru/HDS were reduced using the IRAF software 
(distributed by NOAO\footnote{IRAF is distributed by the National 
Optical Astronomy Observatory, which (AURA) under cooperative 
agreement with the National Science Foundation.}) using common 
reduction procedures of the HDS\footnote{
\url{http://www.naoj.org/Observing/Instruments/HDS/hdsql-e.html}}. 
   We extracted one-dimensional spectra of the comet from the spectral 
images within the aperture of 0''.5 $\times$ 7''.5 to avoid the 
slit-edge regions.
   The wavelength calibration was performed using the Th-Ar lamp 
spectrum and finally, the spectra of comet 21P/GZ are represented 
in the comet's rest frame.
   The flux calibration was performed using bright early-type stars 
near the comet during the observation (see Table \ref{tab:tab1}) 
taking telluric extinction into consideration. The reference stars 
(and their spectral types) are HD 27026 (B9V) and HD 41161 (O8V) for 
the observations on UT 2018 September 5 and 9, and are HD 49643 (B8V) 
and HR 1544 (A1V) on UT 2018 October 3. We assumed the spectrum of 
each reference star as a black-body spectrum at a given temperature 
(effective temperatures of the star). 


   We subtracted the modeled continuum components (represented as the 
sunlight reflected by cometary dust grains) from the reduced spectra to 
extract the emission spectra of comet. The modeled continuum spectrum 
of the comet is computed as a product of the high-resolution solar 
spectrum \citep{Kurucz2005}, the reflectance spectrum of the cometary 
dust grains, and the telluric transmittance spectrum. The reflectance 
spectrum was obtained by dividing the continuum component of the reduced 
spectrum by the solar spectrum. The telluric transmittance spectrum was 
computed using the LBLRTM code \citep{Clough1992} with weather conditions 
at the time of the observations. Finally, the modeled continuum spectrum was 
convolved with the instrumental profile approximated by a Gaussian 
function corresponding to the spectral resolution.

\section{Results and discussion} \label{sec:res}

\subsection{Intensity ratios of the three forbidden oxygen lines}

\begin{figure*}
\includegraphics[angle=-90,width=0.95\textwidth]
                {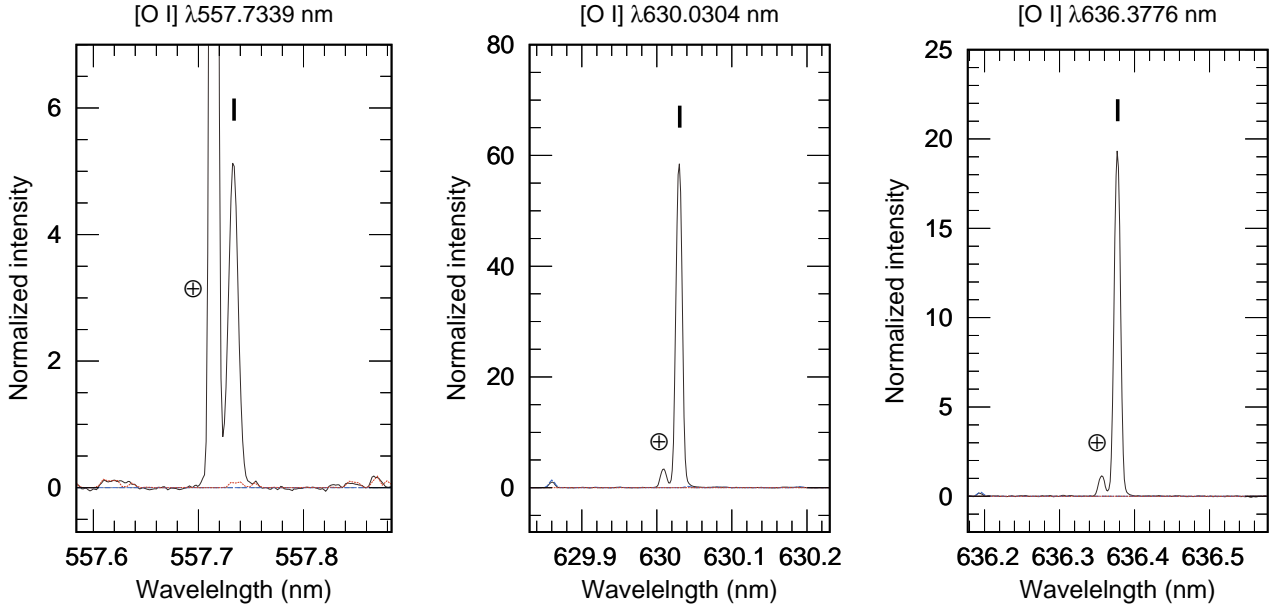}
\caption{
   Three [O I] lines in comet 21P/GZ on UT 2018 October 3. The wavelengths 
are in the cometary rest frame. Vertical tics indicate the [O I] lines 
of 21P/GZ, and the astrological symbols of the Earth (circled "+" marks) 
located at shorter wavelength side of each line originated from telluric 
oxygen. The wavelengths of these telluric [O I] lines were shifted by 
the relative velocity of the comet to the Earth at the observations 
($\dot\Delta =$ 10.49 km s$^{-1}$). The telluric oxygen emission lines 
are clearly separated from the cometary lines. We removed the contamination 
of [O I] emission lines at 557.7 nm by the C$_{2}$ (1--2) Swan band system
(red line). The contamination by C$_{2}$ emission lines with the emission 
line at 557.7 nm is estimated to be 1.7\% $\pm$ 0.3\%. We can ignore 
the contamination of [O I] emission ines at 630.0 and 636.4 nm by the 
NH$_{2}$ (0,8,0) band (blue line).
}
\label{fig:fig1}
\end{figure*}

   In the cometary coma, oxygen atoms are excited to the electronic 
metastable states of $^1$S and $^1$D, and those states emit photons 
at 557.7 and 630.0/636.4 nm as forbidden lines of oxygen, respectively. 
Such excited oxygen atoms can be produced via photodissociation of 
H$_2$O or other oxygen-bearing molecules like CO$_2$ in the coma by 
solar UV radiation. We measured both the intensity and the full width 
at half maximum (FWHM) of the three [O I] lines by fitting them with 
a Gaussian profile for the spectrum taken on UT 2018 October 3 (Figure \ref{fig:fig1}). 
Because the relative velocities between the comet and the Earth 
on UT 2018 September 5 and 9 were too small, the telluric forbidden 
oxygen emission lines overlapped with cometary lines and could not 
be separated. 
   The measured line profile of each [O I] emission line in the observed 
spectrum is a convolution between the intrinsic and instrumental profiles. 
Therefore, the FWHM of the observed emission line ($\rm{FWHM}_{\rm obs}$) 
is expressed by the intrinsic FWHM of the emission line ($\rm{FWHM}_{\rm intr}$) 
and the FWHM of the instrumental profile ($\rm{FWHM}_{\rm inst}$) as 
follows;
\begin{equation}
\rm{FWHM}_{\rm obs} = \sqrt{FWHM_{\rm intr}^2 + FWHM_{\rm inst}^2}, 
\end{equation}
where $\rm{FWHM}_{\rm inst} = 0.00861\pm0.00032$ nm is obtained from 
the telluric nightglow emission lines recorded simultaneously in the 
spectrum of comet 21P/GZ ([O I] at 557.7, 630.0, and 636.4 nm, and Na I 
at 589.0 and 589.6 nm). 

   The [O I] lines at 557.7 nm and 630.0/636.4 nm could be contaminated 
with the emission lines of the C$_{2}$ (1--2) Swan band system and 
the NH$_{2}$ (0,8,0) band, respectively. To measure the intensity of 
these [O I] lines accurately, the contaminations are not negligible 
and must be removed (\citealt{Decock2015, Rousselot2015}, and 
references therein). 
   We measured the emission flux of the [O I] green line at 557.7 nm 
after removing the contamination by C$_{2}$ lines using the C$_{2}$ 
fluorescence excitation model \citep{Shinnaka2010} with a given 
vibrational excitation temperature of 4000 K, which is a typical 
temperature found in comets \citep{Rousselot2012}. 
The contamination 
of the emission line at 557.7 nm by C$_{2}$ emission lines is estimated 
to be 1.7\% $\pm$ 0.3\%.
   For the [OI] lines at 630.0/636.4 nm, we used the synthetic spectrum 
of NH$_{2}$ based on the fluorescence excitation model of NH$_{2}$ 
\citep{Kawakita2000} with an ortho-to-para abundance ratio (OPR) of 
3.31 (see section \ref{sec:OPR}) to subtract the contamination by 
NH$_{2}$. 
The contamination of the [OI] lines at 630.0/636.4 nm by NH$_2$ emission 
is negligible in our spectrum (0.3 $\pm$ 0.1 \% for [O I] at 630.0 nm 
and no NH$_2$ emission lines for [O I] at 636.4 nm). 
Table \ref{tab:tab2} lists the intrinsic intensity and FWHM of each 
[O I] line.

   The resultant intensity ratio of the [O I] red-doublet at 630.0/636.4 
nm was 2.99 $\pm$ 0.04. The derived green-to-red line ratio of [O I] 
(the ratio between the intensity of [O I] at 557.7 nm and the total 
intensity of [O I] red-doublet at 630.0/636.4 nm) was derived to be 
0.074 $\pm$ 0.001, consistent with H$_2$O as the dominant source for 
excited atomic oxygen \citep{CochranCochran2001}. The intrinsic FWHM 
of the green line is wider than that of the red-doublet lines. These 
results are listed in Table \ref{tab:tab2}.

\begin{deluxetable*}{cccccc}
\tablecaption{Results of three [O I] emission lines: Intensity and 
intrinsic FWHM of each line, intensity ratios of red-doublet and 
the green/red lines, and abundance ratio of CO$_{2}$ relative to 
H$_{2}$O. 
\label{tab:tab2}
}
\tablenum{2}
\tablehead{
\multicolumn{3}{c}{Intensity [arbit. units] ($\rm{FWHM}_{\rm intr}$ [km s$^{-1}$] $^{a}$ )} & 
\colhead{$\frac{I_{630.0}}{I_{636.4}}$}  & 
\colhead{$\frac{I_{557.7}}{I_{630.0} + I_{636.4}} ^{b}$} & 
\colhead{$\frac{N_{\rm CO_{2}}}{N_{\rm H_{2}O}}$ [\%] $^{c}$} \\
\cline{1-3}
\colhead{[O I] $\lambda$557.7 nm} & 
\colhead{[O I] $\lambda$630.0 nm} & 
\colhead{[O I] $\lambda$636.4 nm} & 
%
                                  &   
}
\startdata
0.58 $\pm$ 0.01 (2.05$^{+0.16}_{-0.12}$) & 
5.86 $\pm$ 0.06 (0.90$^{+0.22}_{-0.31}$) & 
1.96 $\pm$ 0.02 (0.99$^{+0.20}_{-0.26}$) &
2.99 $\pm$ 0.04   & 
0.0431 $\pm$ 0.0008  & 
(A): 0.9 $\pm$ 0.1  \\
& & & & & 
(B): 11.0  $\pm$ 0.3  \\
\enddata
\tablecomments{\\
$a$: 
   $\rm{FWHM}_{\rm intr}$ [nm] is converted to the most probable velocity 
for the Maxwell-Boltzmann velocity distribution, by using the equation of 
(9) in \citet{Decock2013}. \\
$b$:
   Green-to-red line ratio corrected with a collisional quenching 
factor of 0.58. \\
$c$:
   CO$_{2}$/H$_{2}$O abundance ratio is computed from the green-to-red line 
ratio by the equation (2). The values labeled with (A) and (B) are computed 
with the parameters of cases (A) and (B) in Table \ref{tab:tab_list}, 
respectively (see text).
 }
\end{deluxetable*}

   The intrinsic FWHM of [O I] at 557.7 nm is wider than those of the [O I] 
630.0/636.4 nm lines in contrast with the theoretical prediction for 
the photodissociation of water (as pointed out by \citealt{Cochran2008, 
Decock2013}). 
   \citet{Decock2013} claimed that CO$_2$ is photodissociated with more 
energetic photons than water (that photodissociated mainly by Ly-$\alpha$ 
photon) and therefore the [O I] emission line at 557.7 nm (expected to 
be largely contaminated with O($^1$S) produced from CO$_2$) should be 
wider than the [O I] emission lines at 630.0/636.4 nm (those are mainly 
caused from water and less influenced by CO$_2$). However, to 
discuss the kinetic energies (i.e., velocities) of oxygen atoms produced 
from water and CO$_2$ by photodissociation, we must consider the 
photodissociatoin kinematics of the molecules (e.g., \citealt{Song2014}). 
   An alternative explanation might be possible from the viewpoint of 
lifetimes of excited oxygen atoms since the lifetimes of excited oxygen 
atoms (O($^1$S) and O($^1$D)) are different by a factor of $\sim$100 
(0.79 s and 116 s for 1 au from the Sun). Therefore, the probability for 
collision of O($^1$D) during its lifetime with other molecules (mainly, 
water) in the coma is larger than O($^1$S) by a factor of $\sim$100.
   The meta-stable oxygen atoms (O($^1$S) and O($^1$D)) are chemically 
active and their collisions with water molecules easily produce two 
OH radicals, or the collisions of meta-stable oxygen atoms with water, 
CO$_2$, or CO molecules cause non-radiative transitions to lower 
states \citep{Bhardwaj_Raghuram2012}. Because these collisions of 
O($^1$S) and O($^1$D) with other molecules prevent the [O I] emission, 
only O($^1$D) with smaller velocity differences from the background 
coma molecules, can emit the [O I] emission lines at 630.0/636.4 nm 
while the O($^1$S) atoms with larger velocity differences from 
the background can emit the [O I] emission at 557.7 nm. This may be 
the reason why the [O I] emission line at 557.7 nm is wider than 
the [O I] lines at 630.0/636.4 nm. 

   The obtained intensity ratio of the [O I] red-doublet (630.0/636.4 
nm) is consistent with the ratio of theoretical Einstein's $A$ 
coefficients for the transitions because those transitions have the 
same upper state but different lower states \citep{Galavis1997, 
StoreyZeippen2000}.
   The green-to-red line ratio of [O I] in comet 21P/GZ is similar to
the values previously reported for other comets (\citealt{Capria2010, 
Decock2013, McKay2013, McKay2015, McKay2016}, and references therein), 
supporting the hypothesis that water is the dominant origin of excited 
oxygen atoms generating these three [O I] emission lines in the coma when 
a comet was located closer than $\sim$2.5 au from the Sun. This hypothesis 
is based on the comparison between the observed green-to-red line ratios 
and the ratios of theoretical emission rates of [O I] lines for different 
sources (water, CO, and CO$_{2}$) as claimed by \citet{CochranCochran2001} 
and \citet{Decock2013}.
%
%
   Recent estimates for the [O I] line ratio in the cases of water, CO, and 
CO$_2$ as the source of excited oxygen atoms, are found in the literature 
\citep{Raghuram_Bhardwaj2013, Decock2015, Cessateur2016}. We note that 
no experimentally determined cross-sections for the production of O($^1$S) 
in the photodissociation of water are available \citep{Bhardwaj_Raghuram2012} 
and also note that the yield of O($^1$D) in the photodissociation of 
CO$_2$ is also not reported in the laboratory \citep{HuestisSlanger2006}.
%
%

\begin{figure*}
\centering
\includegraphics[width=0.9\textwidth]
                {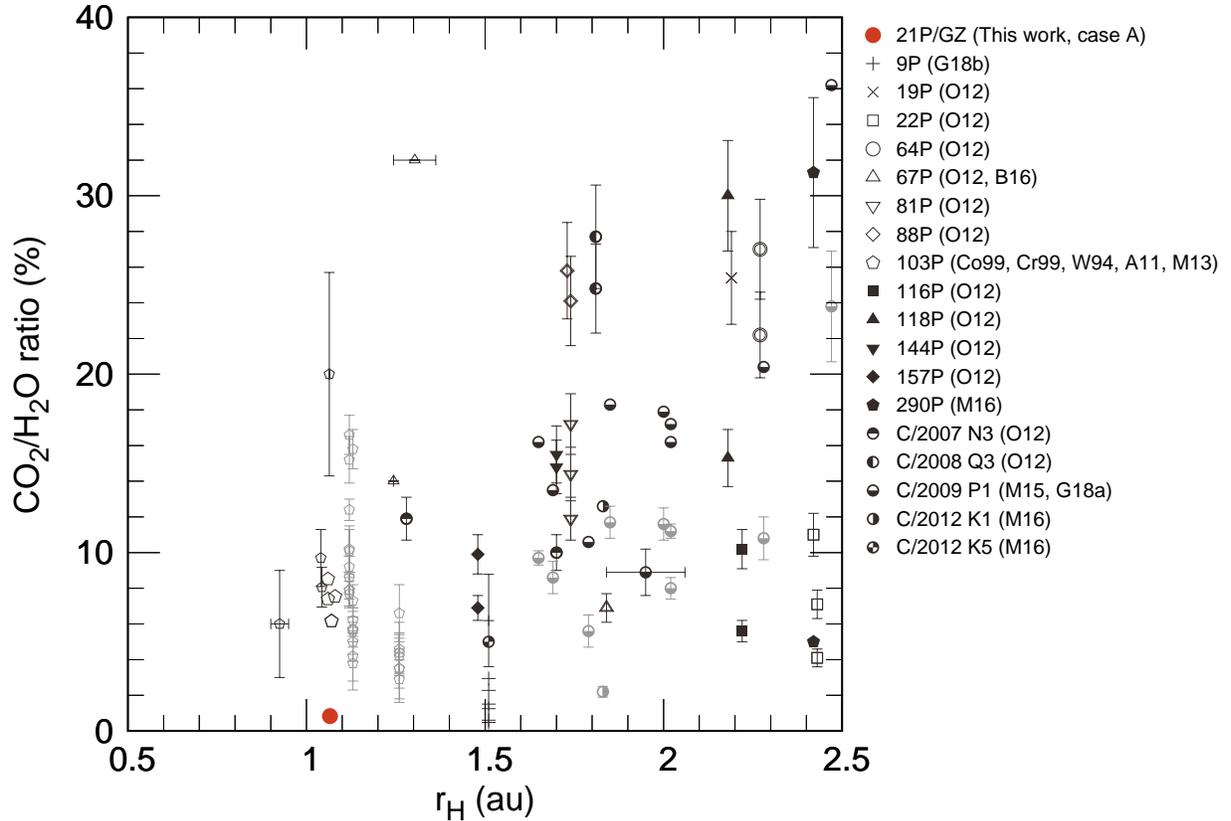}
\caption{CO$_{2}$/H$_{2}$O abundance ratios in comets at observed 
heliocentric distances within 2.5 au. The red (filled-circle) symbol 
is the CO$_{2}$/H$_{2}$O abundance ratio of 21P/GZ estimated from the 
green-to-red line ratio of [O I]. The error bar of comet 21P/GZ is 
smaller than the symbol size. 
   Gray symbols indicate the data in comets estimated from the 
green-to-red line ratios obtained from ground-based facilities in 
consideration of the collisional quenching of O($^1$D) atoms (M13 for 
\citealt{McKay2013}, M15 for \citealt{McKay2015}, and M16 for 
\citealt{McKay2016}), and from the CO Cameron-band observations 
(W94 for \citealt{Weaver1994}).
   Black symbols are the CO$_{2}$/H$_{2}$O abundance ratios in comets 
estimated from the infrared broad-band imaging observations by the 
{\it Spitzer} space telescope (M16 for \citealt{McKay2016}), and 
derived from direct measurements of H$_{2}$O and CO$_{2}$ by 
the {\it ISO} space observatory (Cr99 for \citealt{Crovisier1999} 
and Co99 for \citealt{Colangeli1999}), the {\it AKARI} comet survey 
program (O12 for \citealt{Ootsubo2012}), the {\it Deep Impact} 
spacecraft (A11 for \citealt{AHearn2011}, G18a for \citealt{Gersch2018a} 
and G18b for \citealt{Gersch2018b}), and the {\it Rosetta} spacecraft 
(B16 for \citealt{Bockelee-Morvan2016}), respectively. 
%
\label{fig:fig2}}
\end{figure*}
%

\begin{deluxetable}{lcc}
\tablecaption{Photodissotiation rates (at 1 au) and the branching ratio 
used to convert an observed green-to-red line ratio to a CO$_2$/H$_2$O 
abundance ratio. 
\label{tab:tab_list}}
\tablenum{3}
\tablehead{
\colhead{} & \colhead{Case (A)$^{\rm \, a}$} & \colhead{Case (B)$^{\rm \, b}$} 
}
\startdata
$W_{\rm H_{2}O}^{\rm green}$ & 3.20 $\times$ 10$^{-8}$ s$^{-1}$  & 6.40 $\times$ 10$^{-9}$ s$^{-1}$ \\
$W_{\rm H_{2}O}^{\rm red}$   & 8.00 $\times$ 10$^{-7}$ s$^{-1}$  & 8.44 $\times$ 10$^{-7}$ s$^{-1}$ \\
$W_{\rm CO_{2}}^{\rm green}$ & 7.20 $\times$ 10$^{-7}$ s$^{-1}$  & 3.30 $\times$ 10$^{-7}$ s$^{-1}$ \\
$W_{\rm CO_{2}}^{\rm red}$   & 5.25 $\times$ 10$^{-7}$ s$^{-1}$  & 4.95 $\times$ 10$^{-7}$ s$^{-1}$ \\
$\beta^{\rm green}$       & 0.91 $^{\rm c}$                 & 0.91 $^{\rm c}$  \\
\enddata
\tablecomments{ \\
 a: \citet{Raghuram_Bhardwaj2013}, converted to 1 au. \\
 b: 'McKay2015B' in Table 6 of \citet{McKay2016}.\\
 c: \citet{Slanger2006}.
}
\end{deluxetable}

%
%
%
%

Abundance ratio of CO$_{2}$/H$_{2}$O could be derived from the following 
formula (same as the equation (12) of \citealt{Decock2013});
\begin{equation}
  \frac{N_{\rm CO_{2}}}{N_{\rm H_{2}O}} = 
       \frac{{\rm(G/R)}W_{\rm H_{2}O}^{\rm red} - \beta^{\rm green}W_{\rm H_{2}O}^{\rm green}}
            {\beta^{\rm green}W_{\rm CO_{2}}^{\rm green} - {\rm(G/R)}W_{\rm CO_{2}}^{\rm red}},
\end{equation}
where $N_{\rm X}$ denotes the column density of molecule X, G/R is the 
green-to-red line ratio, $W_{\rm X}^{\rm green}$ and $W_{\rm X}^{\rm red}$ denote 
photodissociation rates of molecule X producing O($^1$S) (green) and O($^1$D)
(red), and $\beta^{\rm green}$ is the branching ratio of the green line at 
557.7 nm for O($^1$S).
Here we assume that only H$_2$O and CO$_2$ are the sources of O($^1$S) and 
O($^1$D) atoms in coma. Note that we might have to consider the production 
of O($^1$S) and O($^1$D) by the photodissociation of O$_2$ molecule in coma. 
In fact, the O$_{2}$ molecule was detected in comet 67P/Churyumov-Gerasimenko 
by the {\it Rosetta}/ROSINA at the first time, and the mean value of the 
local abundance relative to water was reported as 1.8\% $\pm$ 0.4\% 
\citep{Altwegg2019}, which is similar to the value found by the re-analysis 
of {\it in situ} data taken in comet 1P/Halley \citep{Rubin2015}. However, 
based on a recent study by \citet{Cessateur2016}, the contributions of O$_2$ 
to production of O($^1$S) and O($^1$D) are negligible in comparison with 
those of water.

   We applied two parameter sets, the cases (A) and (B) listed in Table 
\ref{tab:tab_list}. The photodissociation rates at 1 au in case (A) are 
basically based on laboratory studies and taken from \citet{Raghuram_Bhardwaj2013} 
while those in case (B) are empirical and taken from 'McKay2015B' in Table 6 
of \citet{McKay2016}. The empirical parameter set successfully reproduced 
the CO$_2$/H$_2$O ratio from [O I] green-to-red line ratio, consistent with 
the CO$_2$/H$_2$O ratio directly measured in infrared for comet C/2009 P1 
(Garradd) and C/2012 K1 (PanSTARRS) although the CO$_2$/H$_2$O ratios derived 
with the parameter set based on laboratory studies are systematically lower 
than those derived with the empirical parameters \citep{McKay2015,McKay2016}.
In order to use the equation (2), we corrected the effect by collisional quenching 
of O($^1$D) atoms in the inner coma on the measured [O I] line intensity, according to 
\citet {McKay2015}. We estimated the fraction of atoms lost to collisional quenching 
based on the Haser model including the quenching of O($^1$D) atoms 
\citep{Morgenthaler2001, Morgenthaler2007}. Because we used small aperture to 
extract the spectrum (the aperture size was 170 km $\times$ 2450 km at the observation),
we assume that H$_2$O molecules are the dominant source of O($^1$D) 
\citep{Raghuram_Bhardwaj2013} and the dominant collision partner \citep{Morgenthaler2001} 
in the inner coma. The water production rate of comet 21P/GZ at the observation was 
assumed to be Q(H$_2$O) $=$ 2.5 $\times$ 10$^{28}$ s$^{-1}$ \citep{Roth2020}.
Furthermore, not only the collisional quenching of O($^1$D) but also that of 
O($^1$S) atoms are considered in our case. The rate coefficients 
for the collisional quenching of O($^1$D) and O($^1$S) by H$_2$O are the same 
as \citet{Decock2015}. The scaling factor for the green-to-red line ratio is 
0.58 in 21P/GZ on UT 2018 October 3. The derived CO$_2$/H$_2$O ratios for both 
parameter sets are listed in Table \ref{tab:tab2}.

   Figure \ref{fig:fig2} shows the CO$_{2}$/H$_{2}$O abundance in comets 
derived by different ways, in addition to our measurement (in case (A)). 
The values plotted in Figure \ref{fig:fig2} are derived from the [O I] 
green-to-red line ratios 
in consideration of the collisional 
quenching of O($^1$D) atoms \citep{McKay2013, McKay2015, McKay2016},
from the CO Cameron-band observations \citep{Weaver1994}, and from the 
direct measurements of H$_{2}$O and CO$_{2}$ by the {\it ISO} space 
observatory \citep{Crovisier1999, Colangeli1999}, the $AKARI$ comet 
survey program \citep{Ootsubo2012}, the {\it Spitzer} space telescope 
\citep{McKay2016}, the {\it Deep Impact} spacecraft \citep{Gersch2018a, 
Gersch2018b, AHearn2011}, and the {\it Rosetta} spacecraft 
\citep{Bockelee-Morvan2016}. 
   Note that the CO$_{2}$/H$_{2}$O ratios derived from the [O I] green-to-red 
line ratios in Figure \ref{fig:fig2} are computed with the parameters of 
the case (A) in Table \ref{tab:tab_list}.

   The comets in Figure \ref{fig:fig2} seem to be classified into two 
groups of comets whose CO$_{2}$/H$_{2}$O ratios are $\sim$10\% and $\sim$25\%, 
and the origin for the bimodal distribution is not clear. Although the 
CO$_2$/H$_2$O ratio might depend on the rotational phase of the cometary 
nucleus \citep{AHearn2011}, comet 21P/GZ is considered to be depleted in 
CO$_2$ compared to water, as shown in Figure \ref{fig:fig2}. 
   The line width of [O I] at 557.7 nm in comet 21P/GZ ($\sim$2.1 km 
s$^{-1}$) (Table \ref{tab:tab2}) is at the lower end of the range of 
intrinsic line widths of [O I] at 557.7 nm in other comets \citep{
Decock2013, Cochran2008, CochranCochran2001}, and this fact is also 
suggestive of the low-CO$_2$ abundance in comet 21P/GZ because CO$_2$ 
might produce O($^1$S) atoms with high-ejection velocities than H$_2$O.
   If we use the empirical parameters taken from \citet{McKay2015} (the 
case (B) in Table \ref{tab:tab_list}), the derived CO$_{2}$/H$_{2}$O ratio 
in comet 21P/GZ is 11.0 $\pm$ 0.3\%, which is higher than that computed 
with the parameters of the case (A) as shown above, but still in the low-CO$_2$ 
group ($\sim$10\%) in Figure \ref{fig:fig2}. 
   Such low-CO$_2$ abundance in comet 21P/GZ could be interpreted as the 
difference in comet-forming regions of these comets or some evolutional 
effects in the inner solar system for Jupiter-family comets.
   \citet{Ootsubo2019} recently proposed that comet 21P/GZ formed in 
the warmer region than other comets, based on their detection of complex 
organics in its low-resolution mid-infrared spectra. 
Our result, low-CO$_2$/H$_2$O in comet 21P/GZ, is consistent their 
hypothesis.

\subsection{Abundance ratio of nuclear spin isomer of the water cation
(H$_{2}$O$^{+}$) and amidogen (NH$_{2}$) \label{sec:OPR}}

\begin{deluxetable}{lcccc}
\tablecaption{Summary of OPRs of NH$_{2}$ and H$_{2}$O$^{+}$ of 21P/GZ. \label{tab:tab_opr}}
\tablenum{4}
\tablehead{
\colhead{ UT Date } & \colhead{Sep 5}  & \colhead{Sep 9} & \colhead{Oct 3} & \colhead{average}
}
\startdata
NH$_{2}$ \\
$\nu_{2}$ = 8  & 3.07 $\pm$ 0.11 & 3.30 $\pm$ 0.09 & 3.26 $\pm$ 0.10 & 3.22 $\pm$ 0.06 \\
$\nu_{2}$ = 9  & 3.34 $\pm$ 0.06 & 3.67 $\pm$ 0.08 & 3.14 $\pm$ 0.09 & 3.39 $\pm$ 0.05 \\
$\nu_{2}$ = 10 & 3.20 $\pm$ 0.19 & 3.14 $\pm$ 0.14 & 3.34 $\pm$ 0.37 & 3.18 $\pm$ 0.11 \\
average        & 3.27 $\pm$ 0.05 & 3.45 $\pm$ 0.06 & 3.20 $\pm$ 0.07 & 3.32 $\pm$ 0.04 \\
\hline
H$_{2}$O$^{+}$ \\ 
$\nu_{2}$ = 10 & 2.73 $\pm$ 0.30 & 2.64 $\pm$ 0.21 & 2.86 $\pm$ 0.18 & 2.76 $\pm$ 0.13 \\
\enddata
\tablecomments{$\nu_{2}$ means (0, $\nu_{2}$, 0) bands of NH$_{2}$ and H$_{2}$O$^{+}$.}
\end{deluxetable}

\begin{figure*}
\includegraphics[angle=0,width=0.90\textwidth]
                {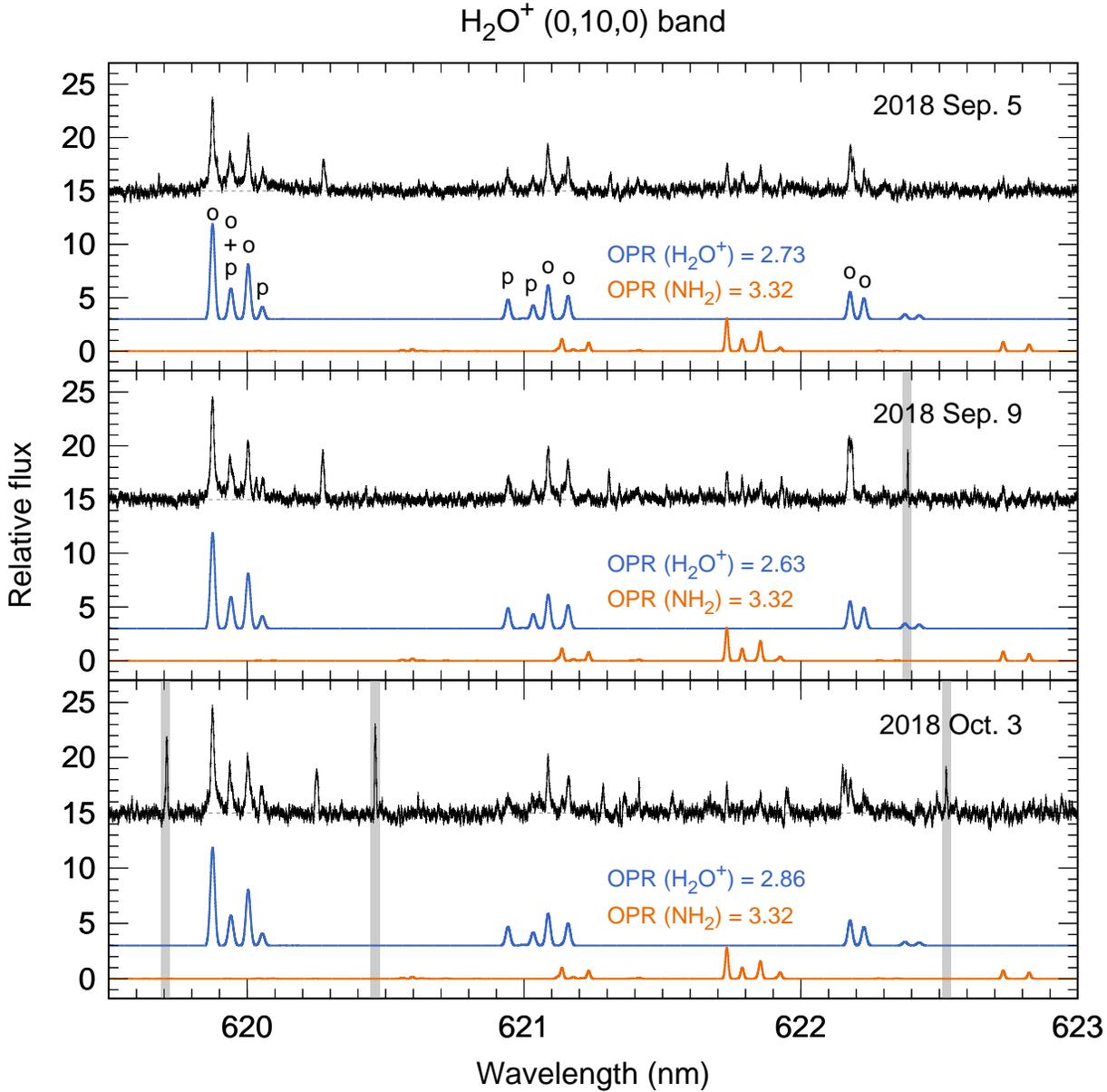}
\caption{
   Comparison between the upper observed (in black, with an offset of 15) 
and lower modeled spectra of H$_{2}$O$^{+}$ (in blue, with an offset of 3) 
as well as NH$_{2}$ (in orange) with an OPR of 3.32 (the average value, see 
Table \ref{tab:tab_opr}) on UT 2018 September 5, 9 and October 3. The ortho- 
and para-lines of H$_{2}$O$^{+}$ ('o' and 'p', respectively) are labeled in 
the top panel. The vertical gray hatches indicate cosmic-ray artifacts.
\label{fig:fig3}
}
\end{figure*}

\begin{figure*}
\includegraphics[angle=0,width=0.90\textwidth]
                {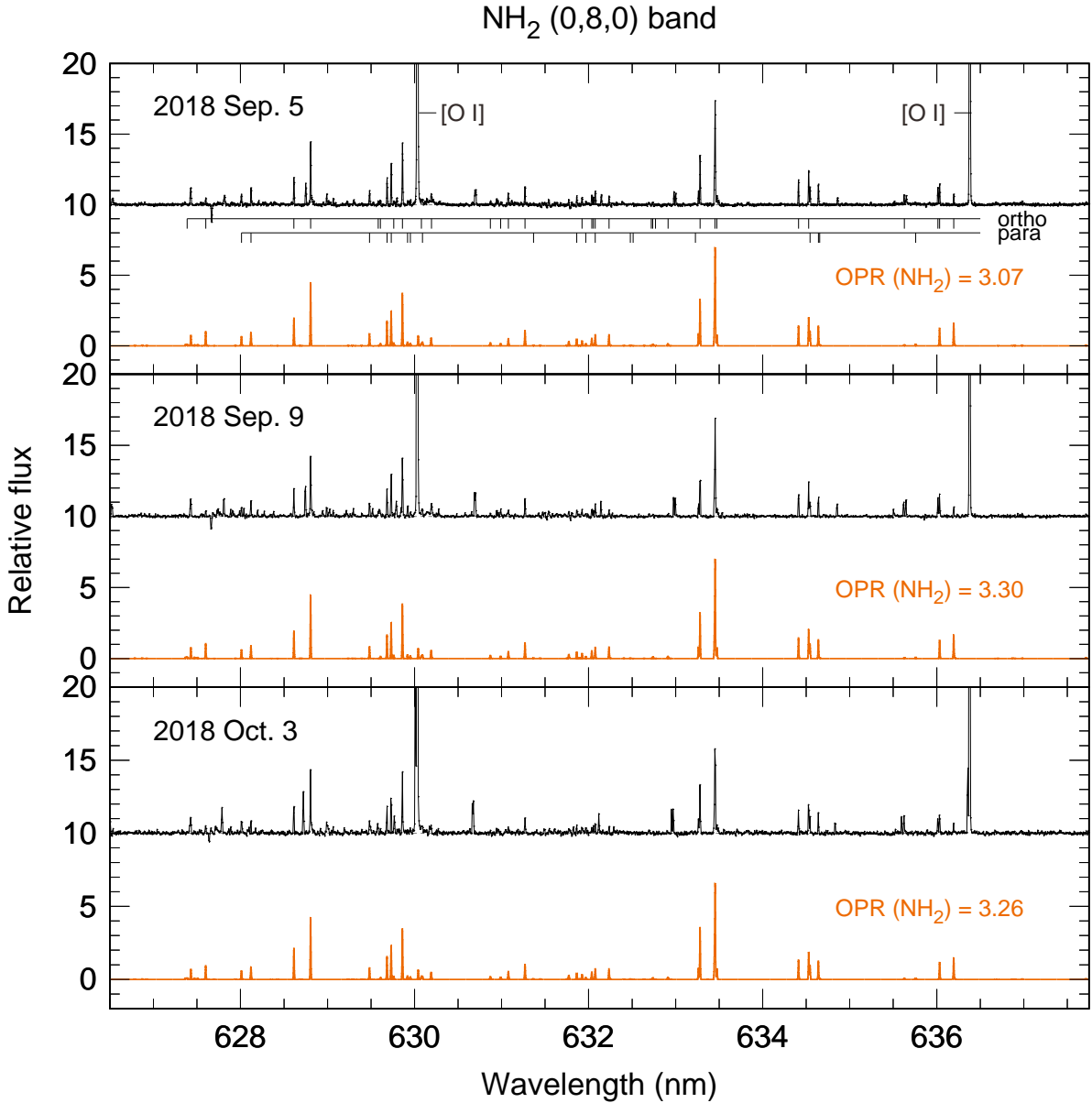}
\caption{
   From top to bottom, comparison between the observed (in black, with 
an offset of 10) and modeled spectra of NH$_{2}$ (0,8,0) band (in orange) 
on UT 2018 September 5, 9, and October 3. The ortho- and para-lines of 
NH$_{2}$ are labeled in these model spectra in the top panel.
   Two strong emission lines at 630.0 nm and 636.4 nm are identified as 
the [O I] lines labeled in the top panel.\label{fig:fig4}
}
\end{figure*}

\begin{figure*}
\includegraphics[angle=0,width=0.90\textwidth]
                {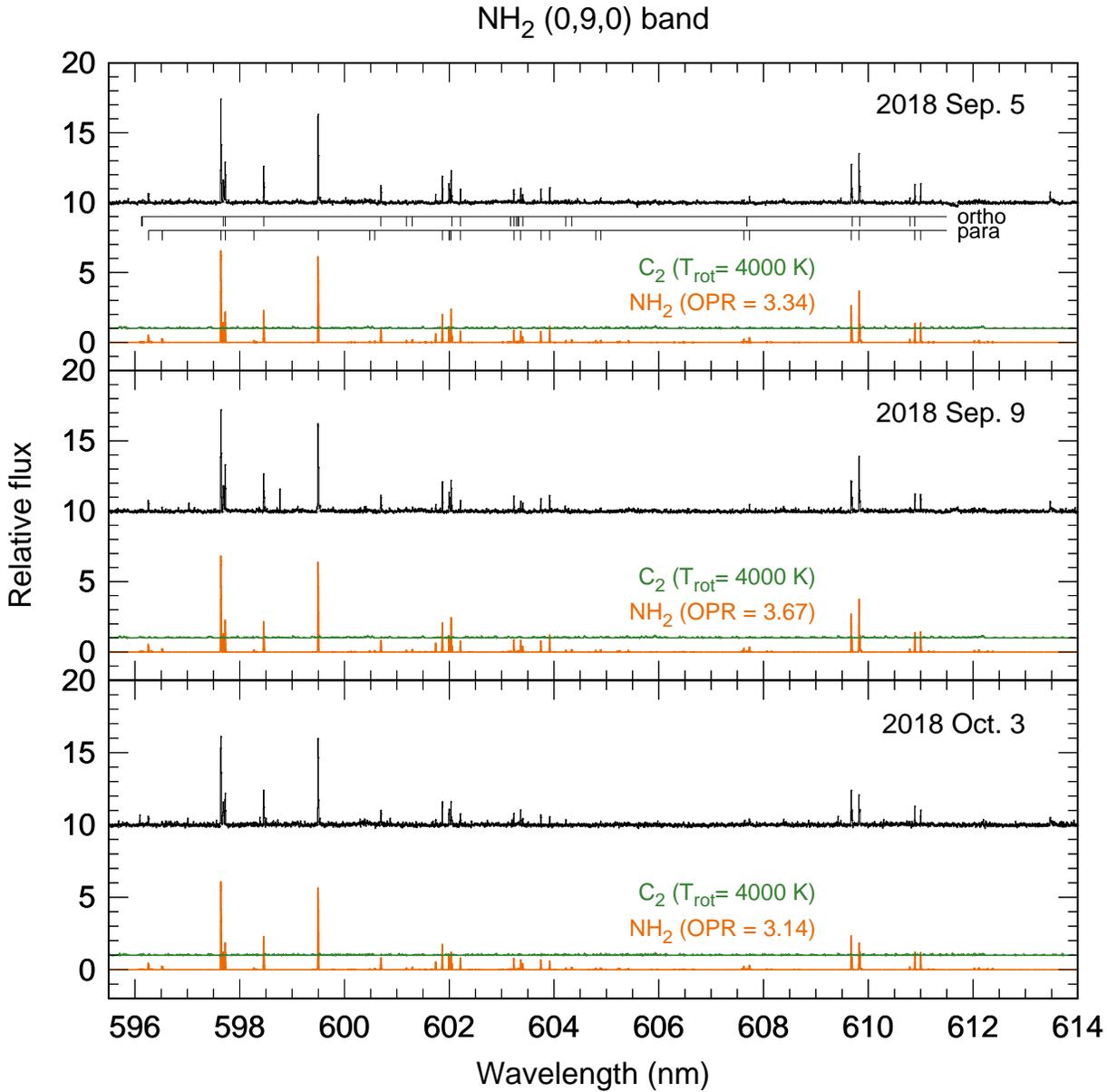}
\caption{
   From top to bottom, comparison between the observed and modeled 
spectra of NH$_{2}$ (0,9,0) band. The modeled spectrum of C$_{2}$ (in 
green, with an offset of 1) is also plotted on the NH$_{2}$ (0,9,0) 
band panels.\label{fig:fig5}
}
\end{figure*}

\begin{figure*}
\includegraphics[angle=0,width=0.90\textwidth]
                 {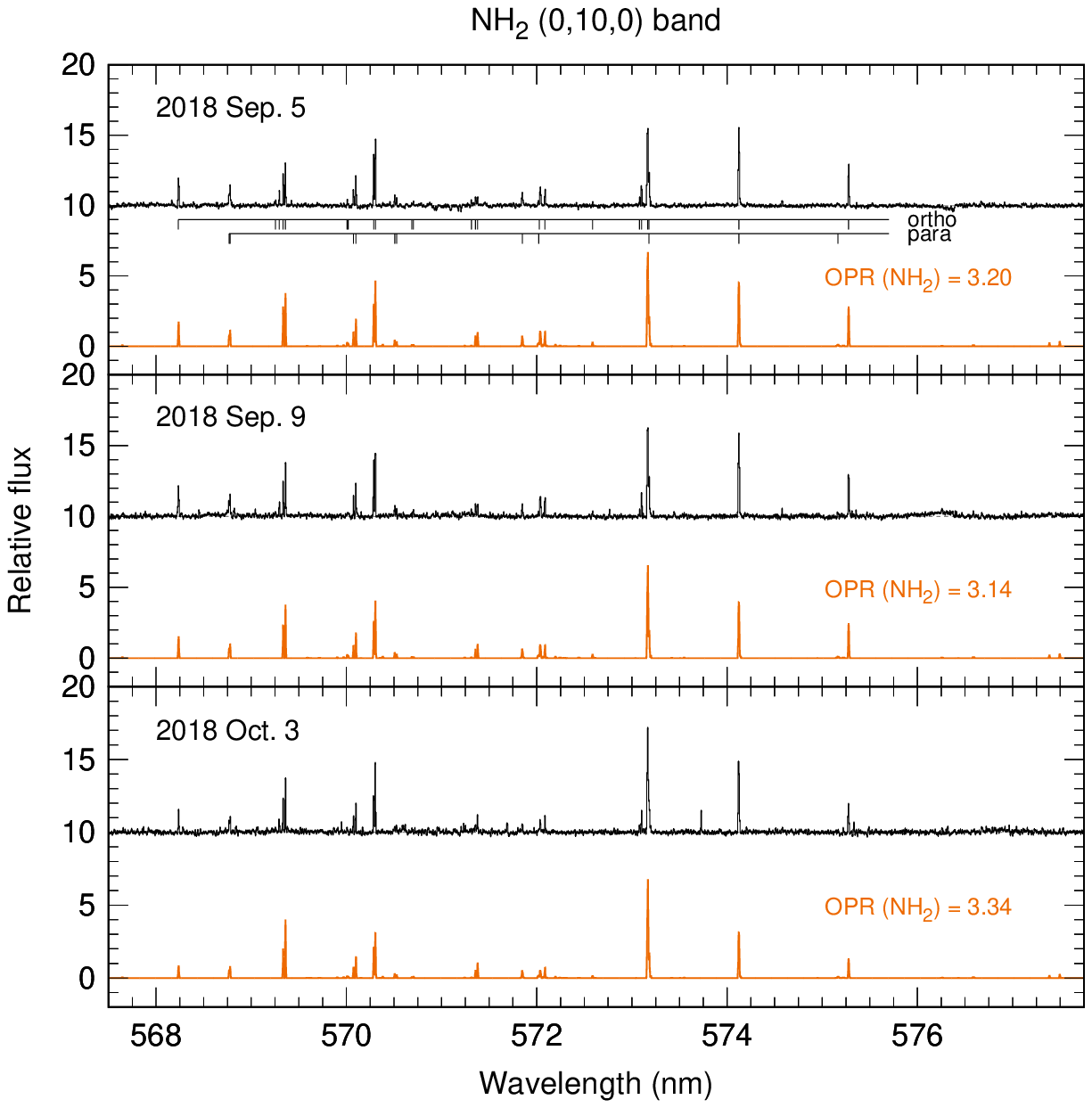}
\caption{
From top to bottom, comparison between the observed and modeled spectra 
of NH$_{2}$ (0,10,0) band. In the NH$_{2}$ (0,10,0) band panels, it 
appears that the model cannot reproduce certain strong lines because 
of the significant line splitting by perturbations described in \citet{Shinnaka2010}. 
The lines affected by the perturbations were simply summed over, and 
the result was compared with the calculation for each unperturbed line.
\label{fig:fig6}
}
\end{figure*}

   We measured the OPRs of water cations (H$_{2}$O$^{+}$) and amidogen 
(NH$_{2}$) in comet 21P/GZ from these rovibronic emissions in the 
high-resolution optical spectrum. Figure \ref{fig:fig3} plots the observed 
and modeled spectra of 21P/GZ around the H$_{2}$O$^{+}$ (0,10,0) band. 
The derived OPRs of H$_{2}$O$^{+}$ and NH$_{2}$ are listed in Table 
\ref{tab:tab_opr}. 
   Figure \ref{fig:fig3} shows that the derived OPR of H$_{2}$O$^{+}$ 
is 2.76 $\pm$ 0.13. Assuming the conservation of total nuclear spin 
through the ionization reaction of water as the sole parent of H$_{2}$O$^{+}$, 
the OPR of H$_{2}$O$^{+}$ is the same as the OPR of water. 
   Indeed, the OPR of water derived from high-resolution near-infrared 
spectra of comet 21P/GZ in its 2005 and 2018 apparitions are OPR = 2.99 
$\pm$ 0.23 \citep{DiSanti2013} and 3.04 $\pm$ 0.12, respectively 
\citep{Faggi2019}. These values are consistent with the OPR of H$_{2}$O$^{+}$ 
obtained here within 3$\sigma$-error interval. 
   The OPR of ammonia is also derived as 1.16 $\pm$ 0.02 based on 
the OPR of NH$_2$ in the comet (see Figures \ref{fig:fig4}, \ref{fig:fig5} 
and \ref{fig:fig6}). Please note that the intensity ratio among bands is 
not correct because we scaled intensity for each plot independently.
   Nuclear spin temperatures of water and ammonia are derived as 
36 +6/$-$3 K from the OPR of H$_{2}$O$^{+}$ and 28 $\pm$ 1 K from the 
OPR of NH$_{2}$ even though the real meaning of the OPRs of water and 
ammonia are unclear.

   Recent laboratory experiments demonstrate that the OPR of water is not 
the memory of its molecular formation \citep{Hama2011, Hama2012, Hama2016, 
Hama2018, Hama_Watanabe2013}. It is likely that this is also for the case 
of ammonia (its OPR is estimated from NH$_{2}$). These laboratory results 
suggest that the OPRs of those molecules are initially the statistical 
weight ratios immediately following the sublimation from nucleus ice. The 
OPRs of cometary volatiles were probably modified by an ortho-para 
conversion process in the inner coma (or other catalyst activities of 
dust crust surfaces of the nucleus) rather than reflected by a temperature 
in the solar nebula 4.6 Ga at the molecular formation. OPRs may be diagnostic 
for the physico-chemical conditions in the inner-most coma or beneath the 
surface.

\subsection{Nitrogen isotopic ratio in NH$_{2}$ \label{sec:14N15N}}
   We also tried to measure the isotopic ratio of nitrogen in NH$_2$ 
($^{14}$N/$^{15}$N) in comet 21P/GZ in the same manner as \citet{Shinnaka_Kawakita2016}. 
Despite clear observation of $^{14}$NH$_2$ (as shown in Figures \ref{fig:fig4} 
and \ref{fig:fig5}), no emission lines of $^{15}$NH$_2$ could be detected 
compared with error levels. The lower limit of $^{14}$N/$^{15}$N in 21P/GZ 
is $>$38 (3$\sigma$) and this value is consistent with the range obtained 
from previous measurements in comets: 135.7 $\pm$ 5.9 \citep{Shinnaka2016}.

\subsection{Formation conditions of comet 21P/Giacobini-Zinner}
   Finally, we discuss the origin of comet 21P/GZ. The depletion of 
simple organic molecules like C$_{2}$H$_{6}$, CH$_{3}$OH, and CO in this 
comet \citep{DiSanti2013} is probably consistent with the depletion of 
carbon-chain molecules such as (C$_{2}$, C$_{3}$) and NH$_{2}$. However, 
these facts do not mean that the comet is depleted in more complex organics 
like PAHs and other hydrocarbons, as observed in 67P/Churyumov-Gerasimenko 
\citep{Altwegg2019}. The observed property of linear polarization produced 
by cometary dust grains indicates the possible existence of complex organic 
matter \citep{Kiselev2000}. 
   Furthermore, \citet{Ootsubo2019} recently reported the detection of 
unidentified IR emission features attributed to complex organic molecules 
such as PAHs (Polycyclic Aromatic Hydrocarbons) in comet 21P/GZ.
Because more complex molecules could form under warmer conditions, comet 
21P/GZ might have formed in a warmer region than where other comets 
formed in the solar nebula. Depletion in highly volatile molecules such 
as C$_{2}$H$_{6}$, CH$_{3}$OH, and CO supports this hypothesis. The 
CO$_2$/H$_2$O ratio in 21P/GZ obtained from our observation is also 
depleted and consistent with the formation under warmer conditions.
   The fluffy and fragile grains of meteoroids of the October Draconids 
meteor shower are also indicative of dust aggregation by organic materials 
acting as glue.

   If comet 21P/GZ formed in the inner region of the solar nebula, the 
dust grains of the comet may contain more abundant crystalline silicates 
(formed in the inner-most coma and transported to the comet-forming region) 
compared to other comets. 
   However, the crystalline-to-amorphous ratio in silicate grains in 
comet 21P/GZ is typical among comets \citep{Ootsubo2019}. Therefore, 
\citet{Ootsubo2019} proposed the hypothesis that comet 21P/GZ formed in 
the circum-planetary disk of giant planets where is warmer than the 
surrounding solar nebula.
\\

\acknowledgments

This paper is based on data collected at the Subaru Telescope, which 
is operated by the National Astronomical Observatory of Japan. We would 
like to thank Editage (www.editage.com) for English language editing.

\software{
IRAF \citep{Tody1986,Tody1993}, LBLRTM code \citep{Clough1992}
}

\end{document}